\documentclass{iopart}

\usepackage{color}
\usepackage{graphicx}
\newcommand{\vect}[1]{\mbox{\boldmath $#1$}}

\newcommand{\changed}[1]{{#1}}

\newcommand{\etabar}{\bar{\eta}}
\newcommand{\rNormalization}{\mathcal{R}}
\newcommand{\figurewidth}{3.4in}

\usepackage{amsmath}
\usepackage{iopams}  

\begin{document}

\title[Optimized quasisymmetric stellarators are consistent with the Garren-Boozer construction]{Optimized quasisymmetric stellarators are consistent with the Garren-Boozer construction}

\author{Matt Landreman}

\address{Institute for Research in Electronics and Applied Physics, University of Maryland, College Park MD 20742, USA}
\ead{mattland@umd.edu}

\begin{abstract}
Most quasisymmetric stellarators to date have been designed by numerically optimizing the plasma boundary shape to minimize symmetry-breaking Fourier modes of the magnetic field strength $B$. At high aspect ratio, a faster approach is to directly construct the plasma shape
from the equations of quasisymmetry near the magnetic axis derived by Garren and Boozer [Phys Fluids B 3, 2805 (1991)]. 
Here we show that the core shape and rotational transform of many optimization-based configurations can be accurately described by this direct-construction approach. 
This consistency supports use of the near-axis construction as an accurate analytical model
for modern stellarator configurations.
\end{abstract}


\section{Introduction}

Magnetic fields with continuous rotational symmetry (axisymmetry) can confine charged particles, since the conserved canonical angular momentum associated with the symmetry strongly constrains the particle orbits;
however a large electric current is required inside the confinement region
which drives instabilities and which is hard to efficiently maintain. Magnetic fields without continuous symmetry (``stellarators'') can confine some particle orbits without any internal current, but as there is no conserved canonical momentum, generally some particle trajectories are not confined. The best features of both the symmetric and non-symmetric cases can be realized in a magnetic field $\vect{B}$ with quasisymmetry, a continuous symmetry of the magnetic field strength $B=|\vect{B}|$ that does not hold for the vector $\vect{B}$.
An outstanding challenge is to understand the landscape of quasisymmetric fields and identify cases that can be realized with practical electromagnetic coils far from the confinement region, enabling good confinement of charged particles without need for an internal current.

Most quasisymmetric configurations to date -- and all quasisymmetric experiments to date -- have been designed using optimization.
An objective function describing the symmetry-breaking Fourier modes of $B$ is minimized, taking the independent variables to
be the shape of a boundary magnetic surface. This approach has been used successfully to find a number of quasisymmetric configurations,
but the method has several drawbacks. The computational cost is high, the fundamental dimensionality of the solution space is unclear, and since results of the optimization depend on the initial condition, one can never prove that all the interesting regions of parameter space have been found.

A complementary approach to finding quasisymmetric configurations is to directly construct the shapes of magnetic surfaces from the underlying equations, either using an expansion about the magnetic axis developed by Garren and Boozer \cite{GB1,GB2,PaperI,PaperII},
 or an expansion about axisymmetry \cite{PlunkHelander}. 
In the former case, although the construction is limited to high aspect ratio, it should be able to describe the core region of any quasisymmetric configuration,
including those with low aspect ratio.
A theorem of existence and uniqueness of solutions in Ref \cite{PaperII} indicates precisely how many
quasisymmetric configurations exist to first order in 
\changed{$r/\rNormalization$, where $r \propto \sqrt{\psi}$ is the effective minor radius,
$2\pi\psi$ is the toroidal flux,
and $\rNormalization$ is a scale length of the magnetic axis, which is best taken to be the minimum radius of curvature \cite{GB1}.}
The solution space is parameterized by the shape of a closed curve (the magnetic axis) which must have nonvanishing curvature everywhere, and three real numbers: $I_2$, $\sigma(0)$, and $\etabar$. The quantities $I_2$ (proportional to the on-axis toroidal current density) and $\sigma(0)$ (related to the angle of the elliptical surface shapes in the $R$-$z$ plane at toroidal angle $\phi=0$) are typically 0. The parameter $\etabar$ (which must be nonzero, and which can be taken to be positive without loss of generality) reflects the magnitude of variation in $B$ on a flux surface:
\begin{align}
B(r,\theta,\varphi) = B_0 \left[ 1 + r \etabar \cos(\theta-N\varphi) + O((r/\rNormalization)^2) \right].
\label{eq:B}
\end{align}
Here, the constant $B_0$ is the field strength on axis,  $\theta$ is the poloidal Boozer angle, $\varphi$ is the toroidal Boozer angle \changed{(which generally differs from the standard toroidal angle $\phi$ by some single-valued function),} and $N$ is a constant integer.
The construction is valid for any plasma pressure, and the pressure does not enter the model to this order.
For any point in the parameter space, 
the construction yields the shape of flux surfaces surrounding the magnetic axis and the rotational transform $\iota$ on the axis.
Each such quasisymmetric configuration can be constructed in $\le 10^{-4}\times$ the computational time
of a single iteration of an optimization \cite{PaperII}, and an optimization typically requires at least hundreds of iterations.
The construction can in principle be carried out to \changed{$O((r/\rNormalization)^2)$} \cite{GB1,GB2}, but the equations are much more cumbersome, and numerical solutions to \changed{this} order have yet to be demonstrated. \changed{In the absence of axisymmetry, quasisymmetry is broken at $O((r/\rNormalization)^3)$ \cite{GB2}.}

The goal of this paper is to show that quasisymmetric configurations in the literature that were obtained by optimization are indeed points in the space of directly constructed configurations. Equivalently, given the axis shape of 
a quasisymmetric configuration obtained by optimization, there exists a value of $\etabar$ such that
the near-axis surface shapes and on-axis rotational transform $\iota$ predicted by the Garren-Boozer construction match those of the optimized configuration. It is not obvious that there would be close agreement; it could have turned out that the small departures from quasisymmetry in optimized configurations led to large departures in the surface shapes compared to the construction. Since however the agreement turns out to be excellent in most cases examined here, including both quasi-axisymmetric and quasi-helically symmetric configurations, we can be confident that the direct construction does include (the cores of) the configurations that will be interesting in practice. This knowledge is valuable because it means the uniqueness theorem  \cite{PaperII} for the near-axis construction
is applicable in practice, despite the presence of small departures from quasisymmetry in practical designs.
Then the direct construction can be used to survey the entire parameter space of 
quasisymmetric stellarator cores, and we can be confident that no interesting regions of parameter space will be missed. This is in contrast to conventional optimization-based stellarator design, where one can never be certain that a local optimum which has been found is the global optimum.

The study here includes a wide range of configurations, developed by several independent research teams using different optimization codes.
The configurations span values of aspect ratio from $2.6 - 12$, numbers of field periods $n_{fp} = 2-6$, and normalized plasma pressure $\beta$ in the range $0 - 4\%$.
In eight of the ten configurations considered we find an excellent fit to the construction.
In the other two cases (ARIES-CS and CFQS), the construction predicts somewhat higher elongation, but the agreement in the angular orientation of the ellipse at each toroidal angle is still  good. We will show these two cases can be explained by a relatively larger magnitude of symmetry-breaking near the axis in the optimized configurations.
For all the optimized configurations, we will only attempt to show agreement with the Garren-Boozer construction close to the magnetic axis,
since the construction is based on a near-axis expansion, so from regularity considerations only magnetic surface shapes with elliptical cross sections can be represented.
Quasi-poloidal symmetry will not be considered here, since it is always broken at \changed{$O(r/\rNormalization)$}.


\section{Method and Results}

We now explain in detail how a directly constructed quasisymmetric configuration is fit to a given optimized configuration.
The equations for the direct construction approach are summarized in section 2 of \cite{PaperII}, and need not be repeated here.
The optimized configuration is represented using the VMEC code \cite{VMEC1983}, and the magnetic axis shape from the optimized configuration is adopted for the construction.
All the optimized configurations considered here have vanishing toroidal current density on the magnetic axis, so we take $I_2=0$.
Also, all the optimized configurations considered here possess stellarator symmetry, so we take $\sigma(0)=0$.
Given each flux surface from the optimized configuration, the corresponding effective minor radius parameter $r$ for the construction can be identified as $r=\sqrt{2 |\psi| / B_0}$, where 
$B_0$ is the  field strength averaged along the axis.

\changed{
Thus, to fit the construction to an optimized configuration, the only available freedom is fitting the single number $\bar{\eta}$. From (\ref{eq:B}) we see that $\bar{\eta}$ controls the magnitude of variation of $B$ over flux surfaces, and the meaning of $\bar{\eta}$ can be further understood by considering axisymmetry with up-down symmetry. In this case eq (B4) of \cite{PaperI} with (2.5)-(2.6) of \cite{PaperII} shows the flux surface elongation is $\max(\bar{\eta} R_0, \, 1/(\bar{\eta} R_0))$, where $R_0$ is the major radius of the axis, and values of $\bar{\eta} R_0$ greater (less) than 1 give wide (tall) ellipses. Thus, $\bar{\eta}$ essentially controls the elongation. In quasisymmetry, the elongation is no longer precisely given by the same function of $\bar{\eta}$, but the relation between $\bar{\eta}$ and elongation remains qualitatively similar, and $\bar{\eta}$ can be expected to be within an order-unity multiplicative factor of $1/R_0$.
}

To fit $\etabar$ for an optimized configuration, 
we numerically minimize an objective function $f(\etabar)$, defined as follows.
For a given value of $\etabar$, the Garren-Boozer construction is carried out, and the $R$ and $z$ coordinates of the surface are 
evaluated on a uniform tensor product grid $(\vartheta_j,\phi_k)$ (where $j=1\ldots N_\vartheta$, $k=1\ldots N_\phi$) in the coordinates $(\vartheta,\phi)$, where $\phi$ is the  toroidal angle (the standard azimuthal angle of cylindrical coordinates) and $\vartheta$ is \changed{a poloidal angle that will be discussed shortly.}
 In the same way, $R(\vartheta_j,\phi_k)$ and $z(\vartheta_j,\phi_k)$ are evaluated for the chosen surface of the optimized configuration.
We then define
\begin{align}
\label{eq:residual}
f(\etabar) = & \left[ \frac{1}{N_\vartheta N_\phi}
\sum_{j=1}^{N_\vartheta} \sum_{k=1}^{N_\phi} \left( [R_{opt}(\vartheta_j,\phi_k)-R_{dc}(\vartheta_j,\phi_k)]^2
\right.\right.\\
& \hspace{1.0in}\left.\vphantom{\sum_{k=1}^{N_\phi}}\left.
+[z_{opt}(\vartheta_j,\phi_k)-z_{dc}(\vartheta_j,\phi_k)]^2 \right) \right]^{1/2},
\nonumber
\end{align}
where the subscripts $opt$ and $dc$ refer to the optimized and direct-construction surfaces. Minimizing $f(\etabar)$ yields the best-fit value of $\etabar$.

\changed{
In the above procedure, a challenge is that the original poloidal angle in the direct-construction model and the poloidal angle in the optimized configuration generally differ, so the maps from poloidal angle to cylindrical coordinates $(R,z)$ may differ even if the surface shapes coincide.
To avoid this issue, the fit should be done using a poloidal angle $\vartheta$ that can be defined in the same way for the two configurations. 
One suitable choice of $\vartheta$ is atan$((z-z_0)/(R-R_0))$, where $(R_0,z_0)$ are the cylindrical coordinates of the magnetic axis. For elliptical surfaces, this angle results in the sum (\ref{eq:residual}) being weighted towards the small-diameter regions of the ellipses, so a second definition of $\vartheta$ was considered, in which the arclength in the $R$-$z$ plane is uniform at each $\phi$. For this angle, the $\vartheta=0$ direction is chosen at each $\phi$ to point horizontally outward from the magnetic axis. It was found that the two definitions of $\vartheta$ yield nearly indistinguishable results, with the best-fit values of $\etabar$ differing by $<6$\% (and by $<1.4$\% for 8 of the 10 configurations examined). Results using the uniform-arclength $\vartheta$ will be shown here.
}

\begin{figure}
  \centering
  \includegraphics[width=\figurewidth]{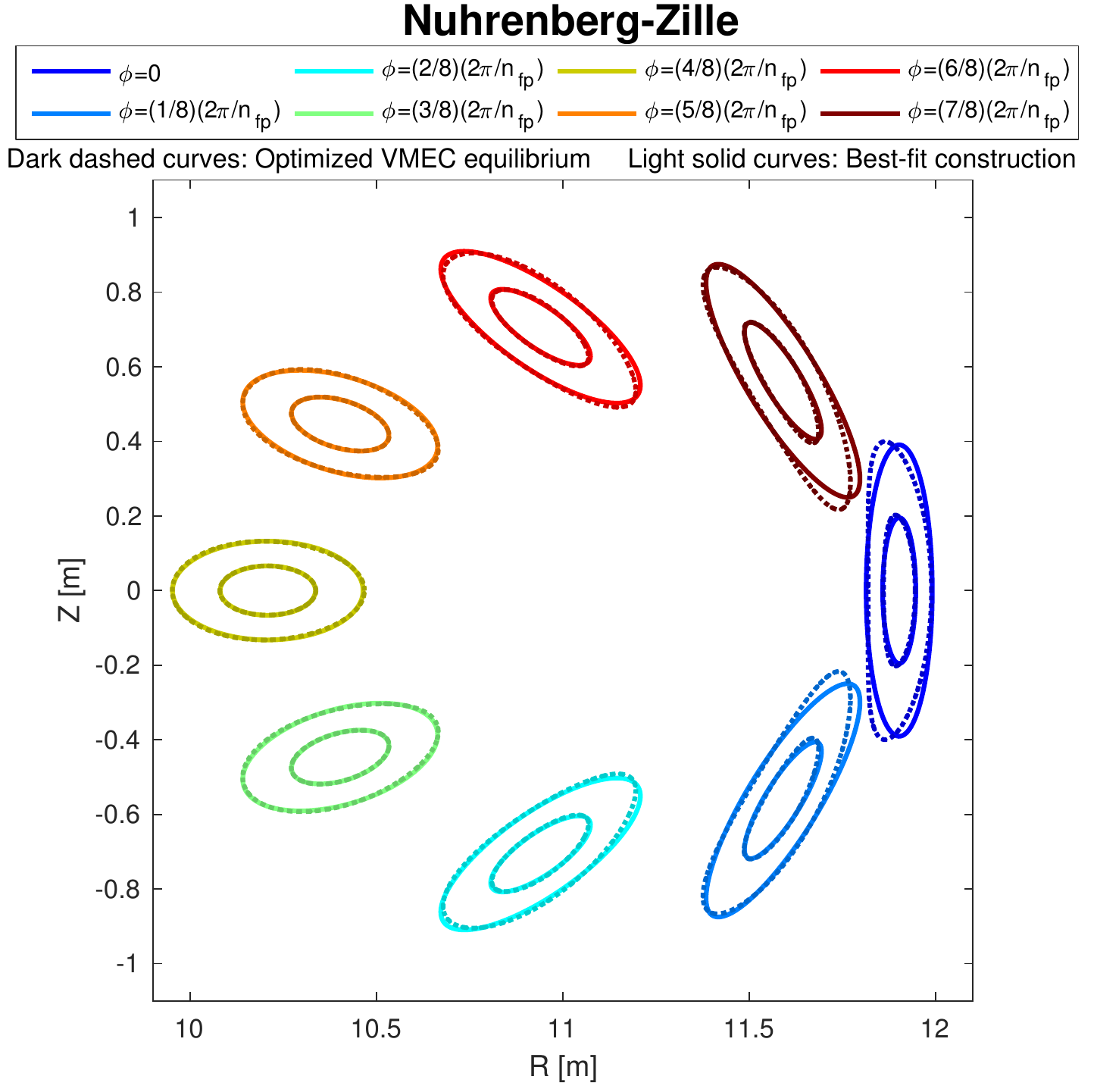}
  \caption{
Shapes of the magnetic surfaces at $\rho=\sqrt{s}=0.1$ and $0.2$ for the configuration of 
figure 1 of \cite{NuhrenbergZille} (dark dashed curves) at 8 toroidal locations, and the same surfaces of the corresponding direct-construction quasisymmetric configuration (solid curves).
  }
\label{fig:NuhrenbergZille}
\end{figure}

\begin{figure}
  \centering
  \includegraphics[width=\figurewidth]{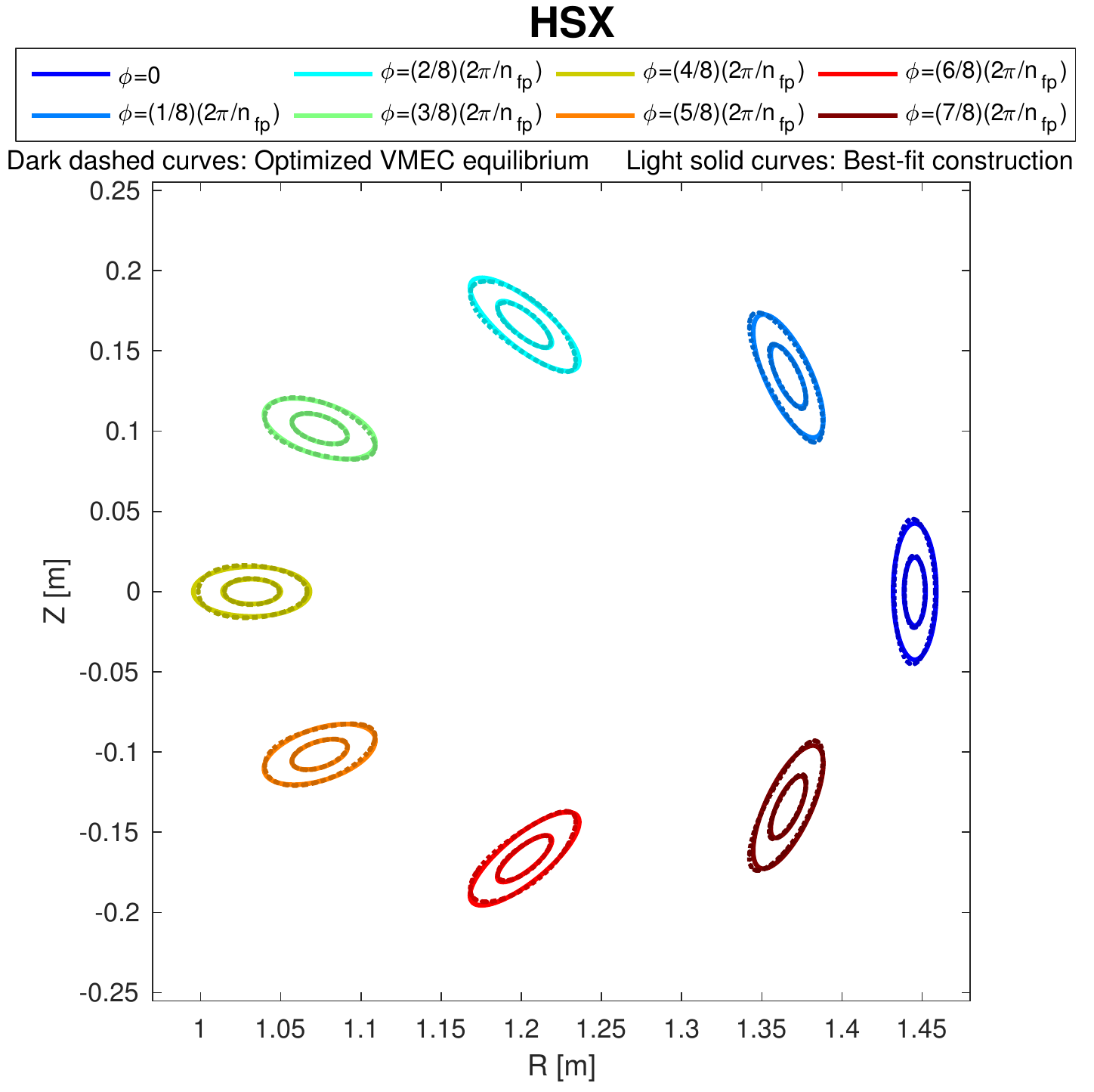}
  \caption{
Shapes of the magnetic surfaces at $\rho=\sqrt{s}=0.1$ and $0.2$ for HSX \cite{HSX} (dark dashed curves) at 8 toroidal locations, and the same surfaces of the corresponding direct-construction quasisymmetric configuration (solid curves).
  }
\label{fig:HSX}
\end{figure}

\begin{figure}
  \centering
  \includegraphics[width=\figurewidth]{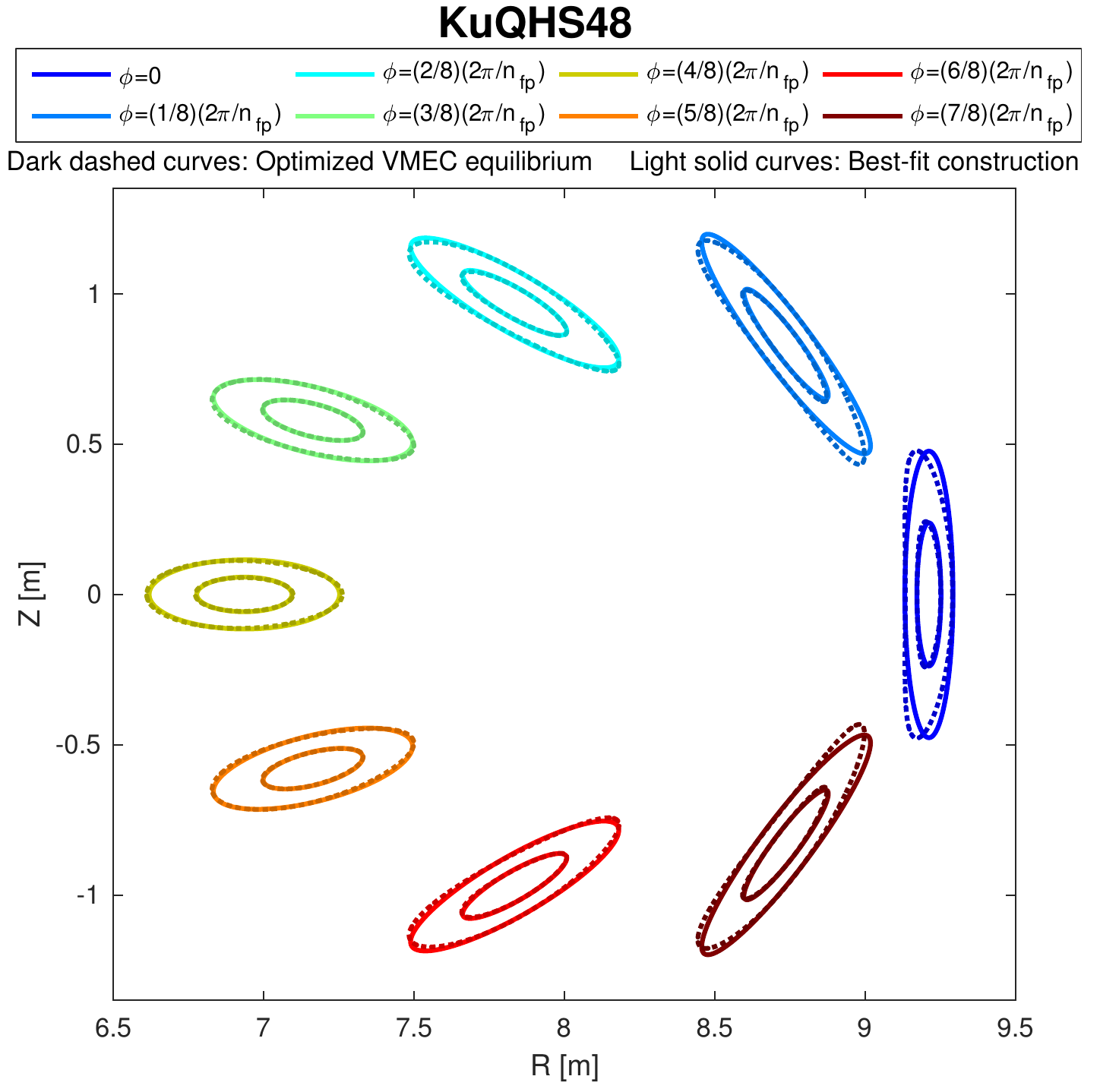}
  \caption{
Shapes of the magnetic surfaces at $\rho=\sqrt{s}=0.1$ and $0.2$ for the quasi-helically symmetric
configuration of section 2 of \cite{KuBoozerQHS} (dark dashed curves) at 8 toroidal locations, and the same surfaces of the corresponding direct-construction quasisymmetric configuration (solid curves).
  }
\label{fig:KuQHS48}
\end{figure}

\begin{figure}
  \centering
  \includegraphics[width=\figurewidth]{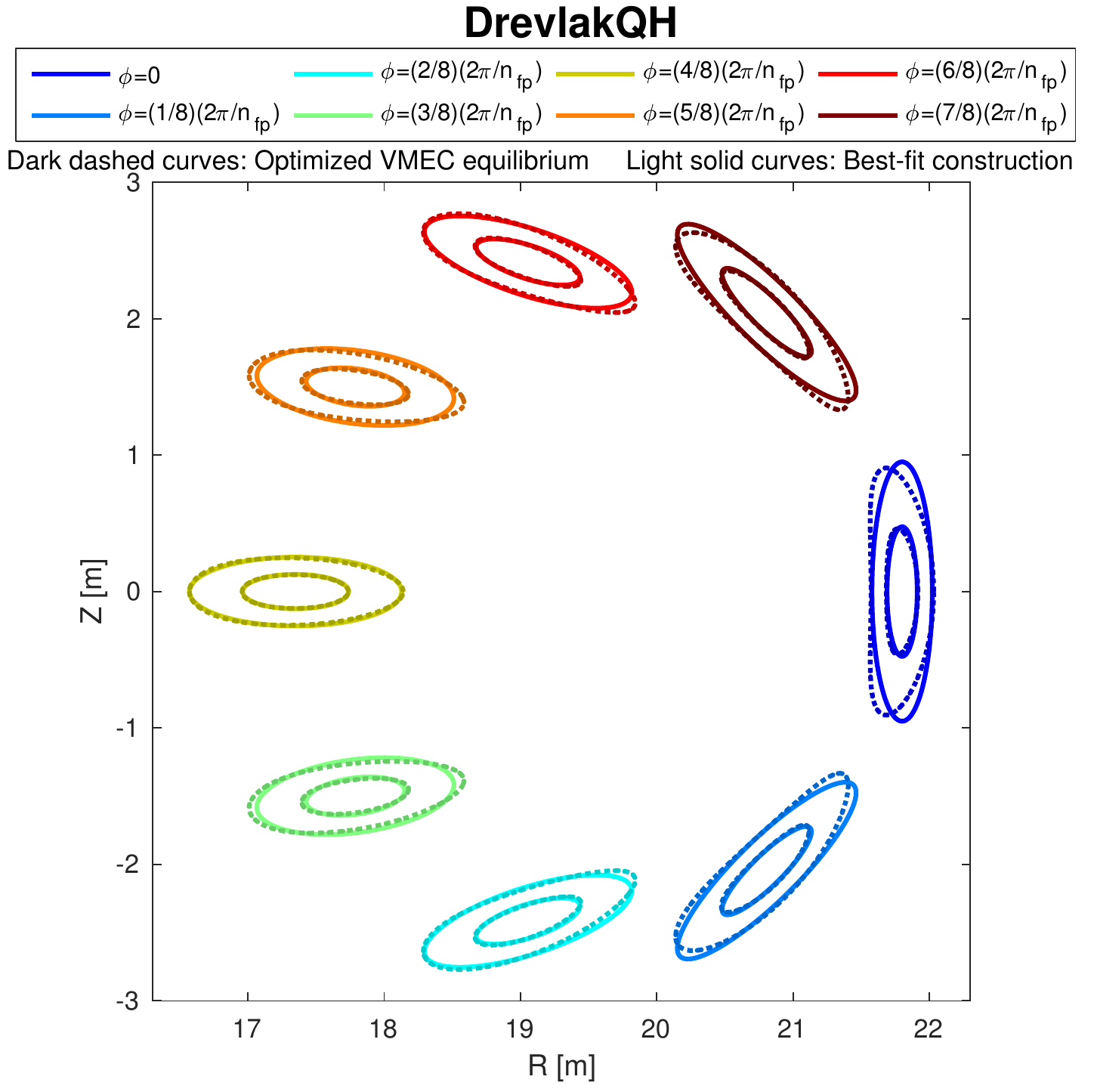}
  \caption{
Shapes of the magnetic surfaces at $\rho=\sqrt{s}=0.1$ and $0.2$ for the quasi-helically symmetric configuration of Ref \cite{DrevlakKyoto} (dark dashed curves) at 8 toroidal locations, and the same surfaces of the corresponding direct-construction quasisymmetric configuration (solid curves).
  }
\label{fig:DrevlakQH}
\end{figure}

\begin{figure}
  \centering
  \includegraphics[width=\figurewidth]{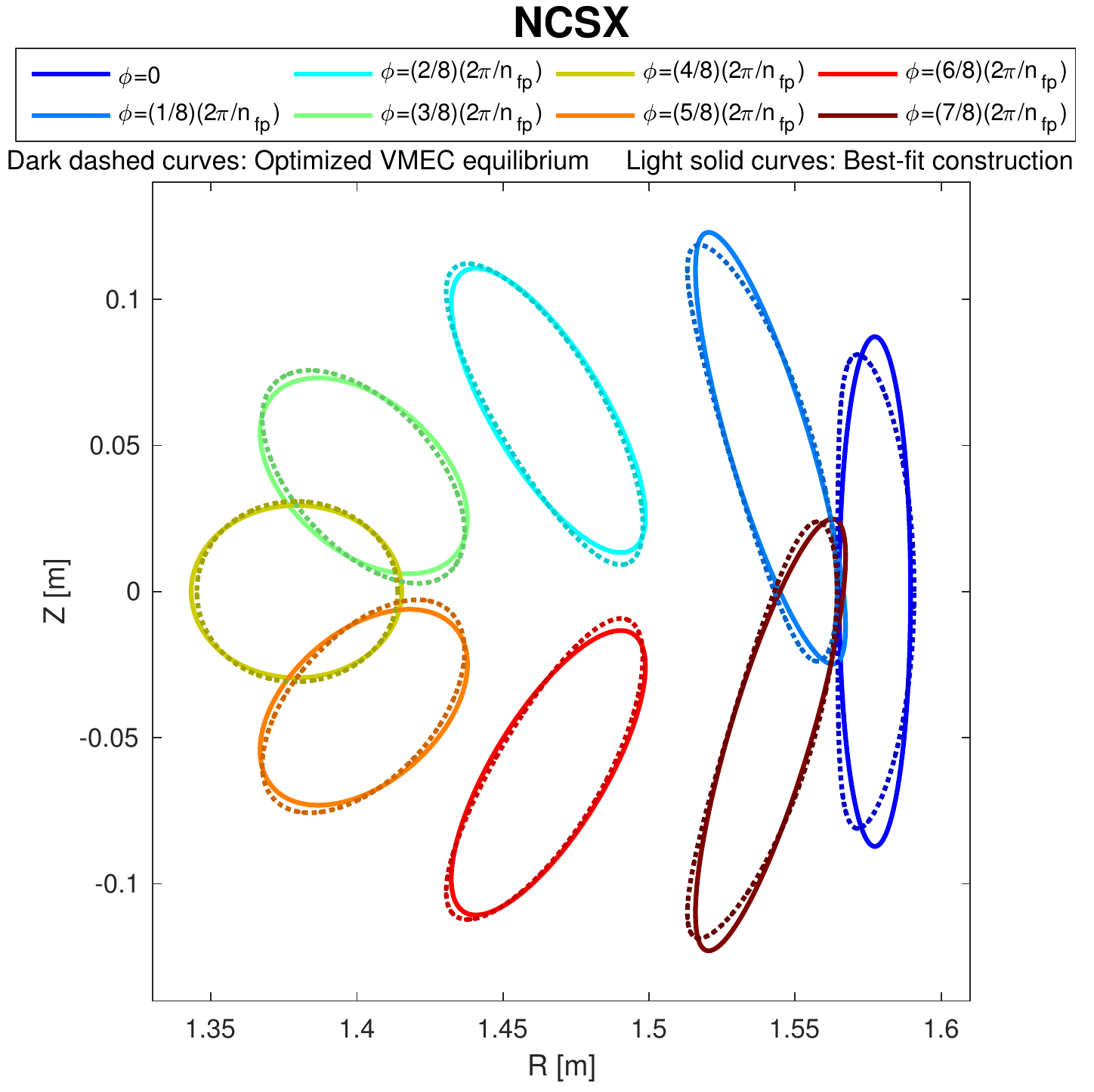}
  \caption{
Shapes of the magnetic surface at $\rho=\sqrt{s}=0.1$ 
for NCSX \cite{NCSX} (dark dashed curves) at 8 toroidal locations, and the same surface of the corresponding direct-construction quasisymmetric configuration (solid curves).
  }
\label{fig:NCSX}
\end{figure}

\begin{figure}
  \centering
  \includegraphics[width=\figurewidth]{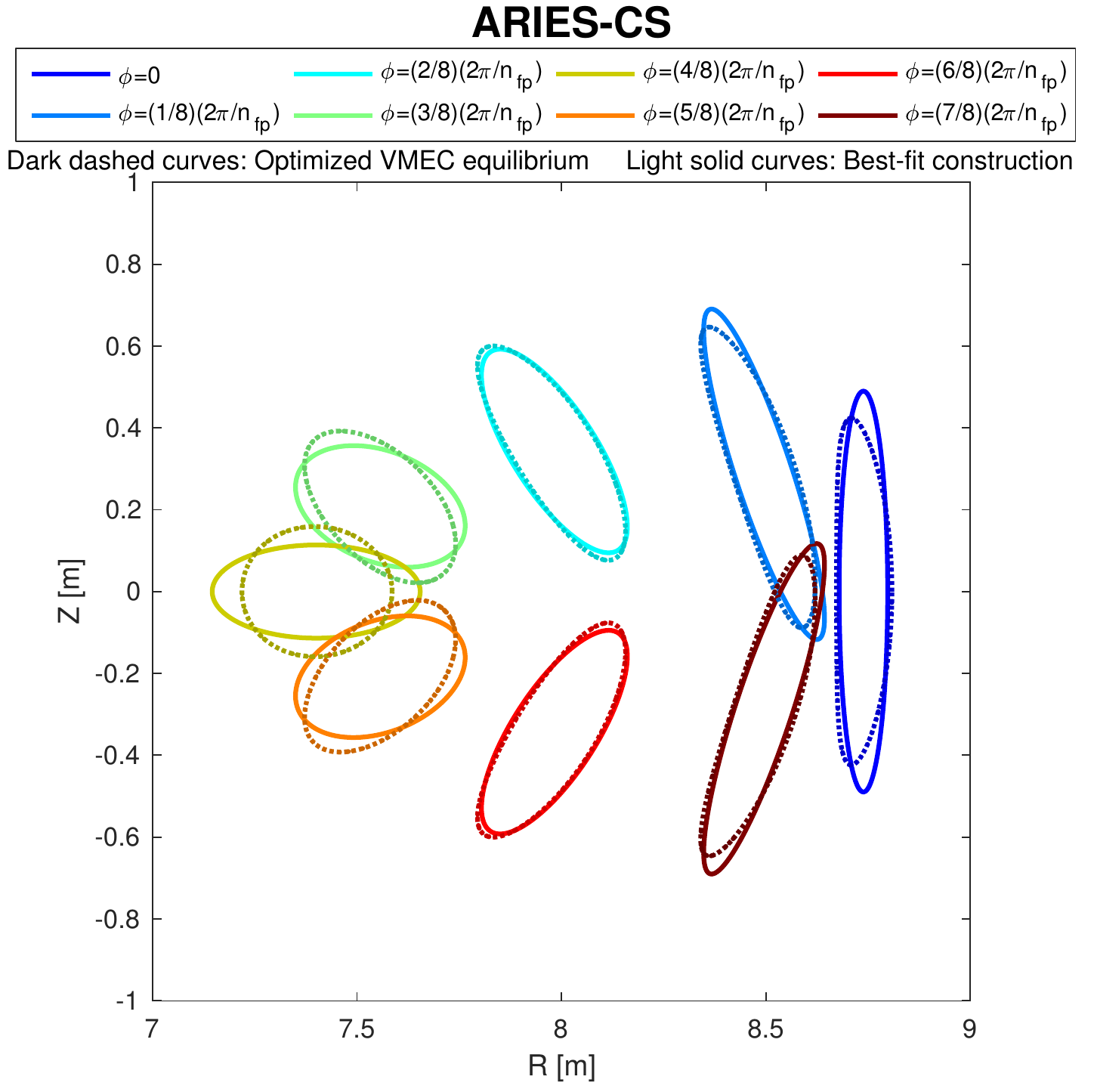}
  \caption{
Shapes of the magnetic surface at $\rho=\sqrt{s}=0.1$ 
for ARIES-CS \cite{ARIESCS} (dark dashed curves) at 8 toroidal locations, and the same surface of the corresponding direct-construction quasisymmetric configuration (solid curves).
  }
\label{fig:ARIESCS}
\end{figure}

\begin{figure}
  \centering
  \includegraphics[width=\figurewidth]{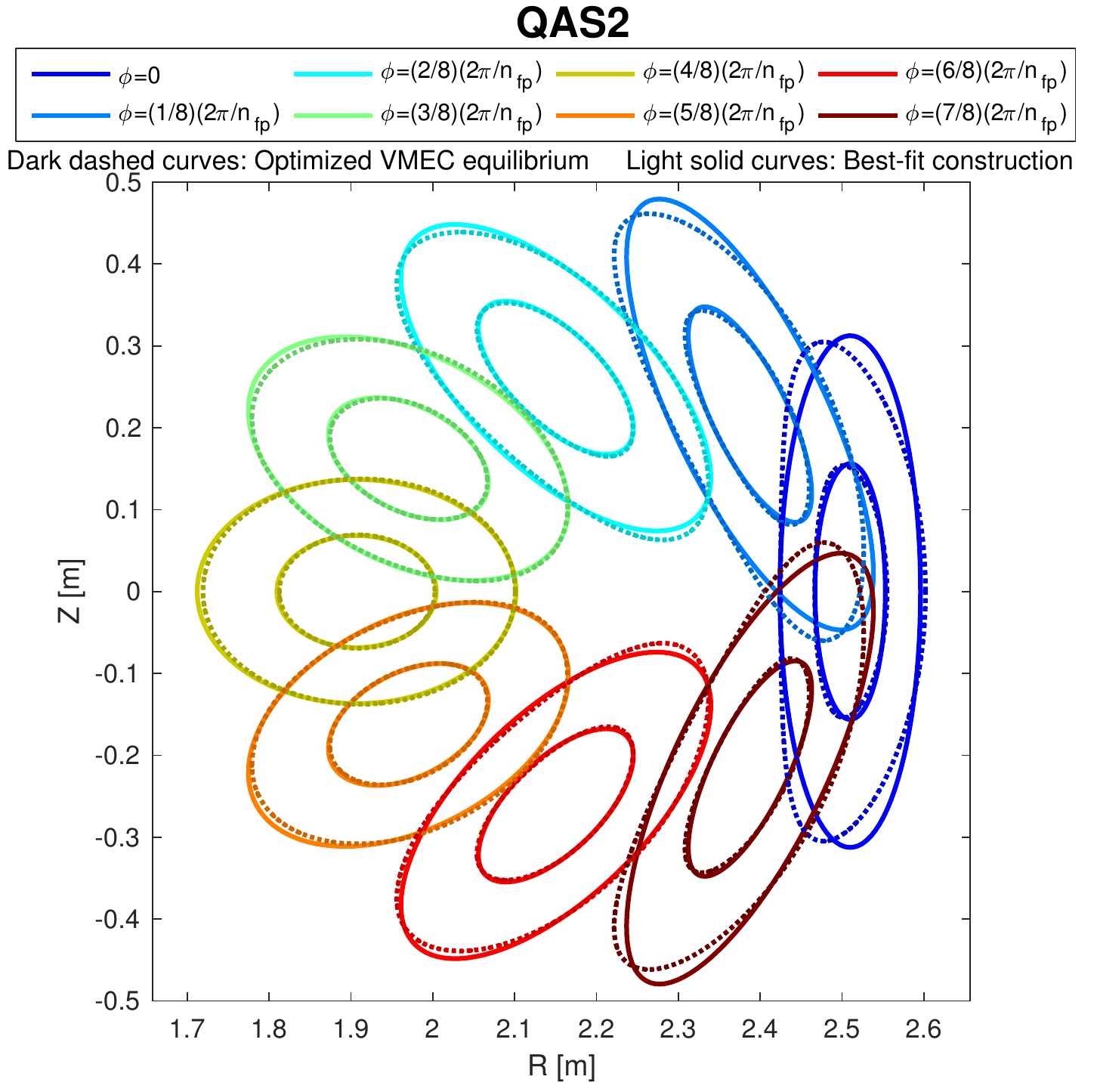}
  \caption{
Shapes of the magnetic surfaces at $\rho=\sqrt{s}=0.1$ and $0.2$ for QAS2  \cite{GarabedianPNAS,GarabedianNIST} (dark dashed curves) at 8 toroidal locations, and the same surfaces of the corresponding direct-construction quasisymmetric configuration (solid curves).
  }
\label{fig:QAS2}
\end{figure}

\begin{figure}
  \centering
  \includegraphics[width=\figurewidth]{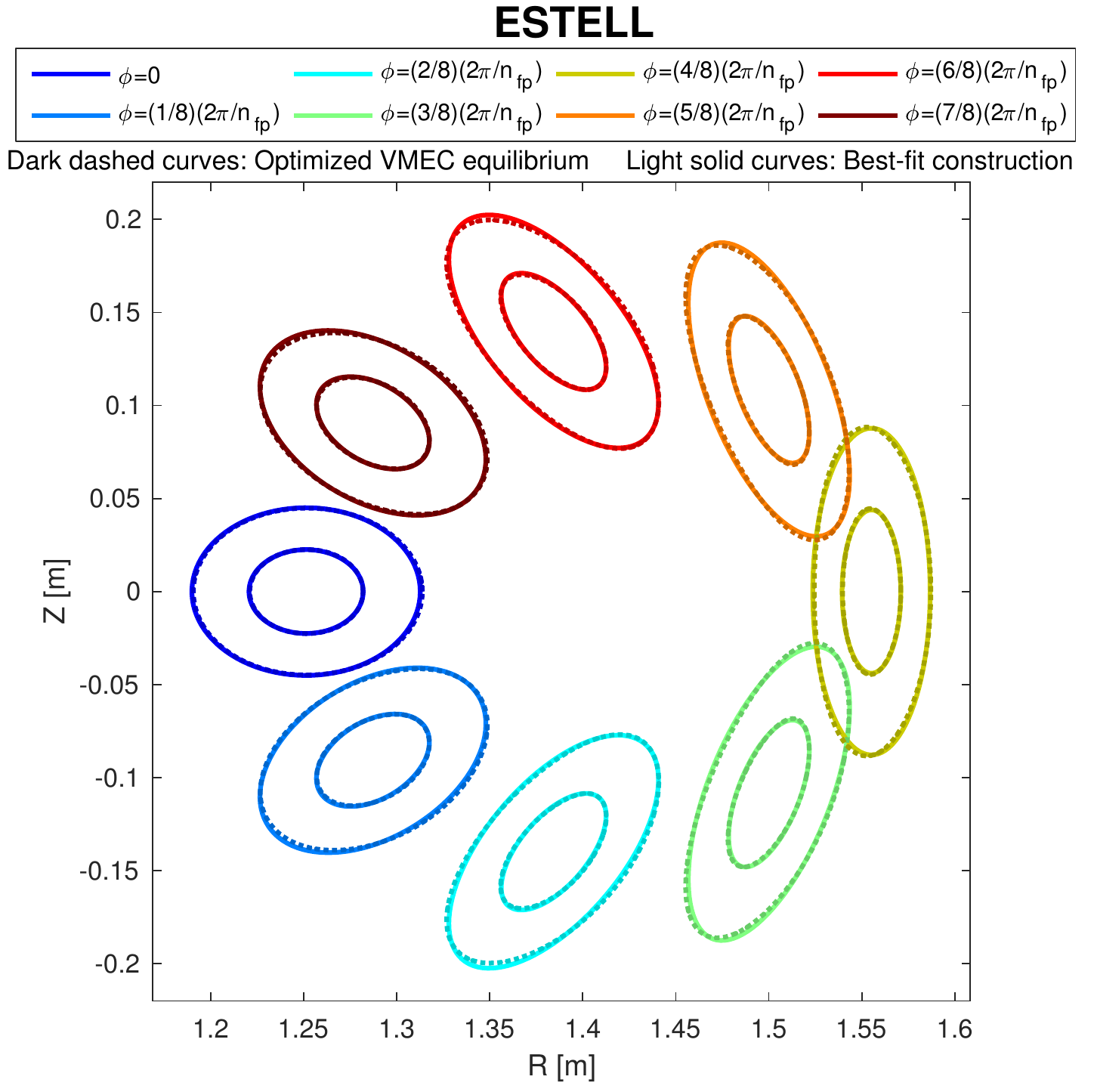}
  \caption{
Shapes of the magnetic surfaces at $\rho=\sqrt{s}=0.1$ and $0.2$ for ESTELL \cite{ESTELL,ROSE} (dark dashed curves) at 8 toroidal locations, and the same surfaces of the corresponding direct-construction quasisymmetric configuration (solid curves).
  }
\label{fig:ESTELL}
\end{figure}

\begin{figure}
  \centering
  \includegraphics[width=\figurewidth]{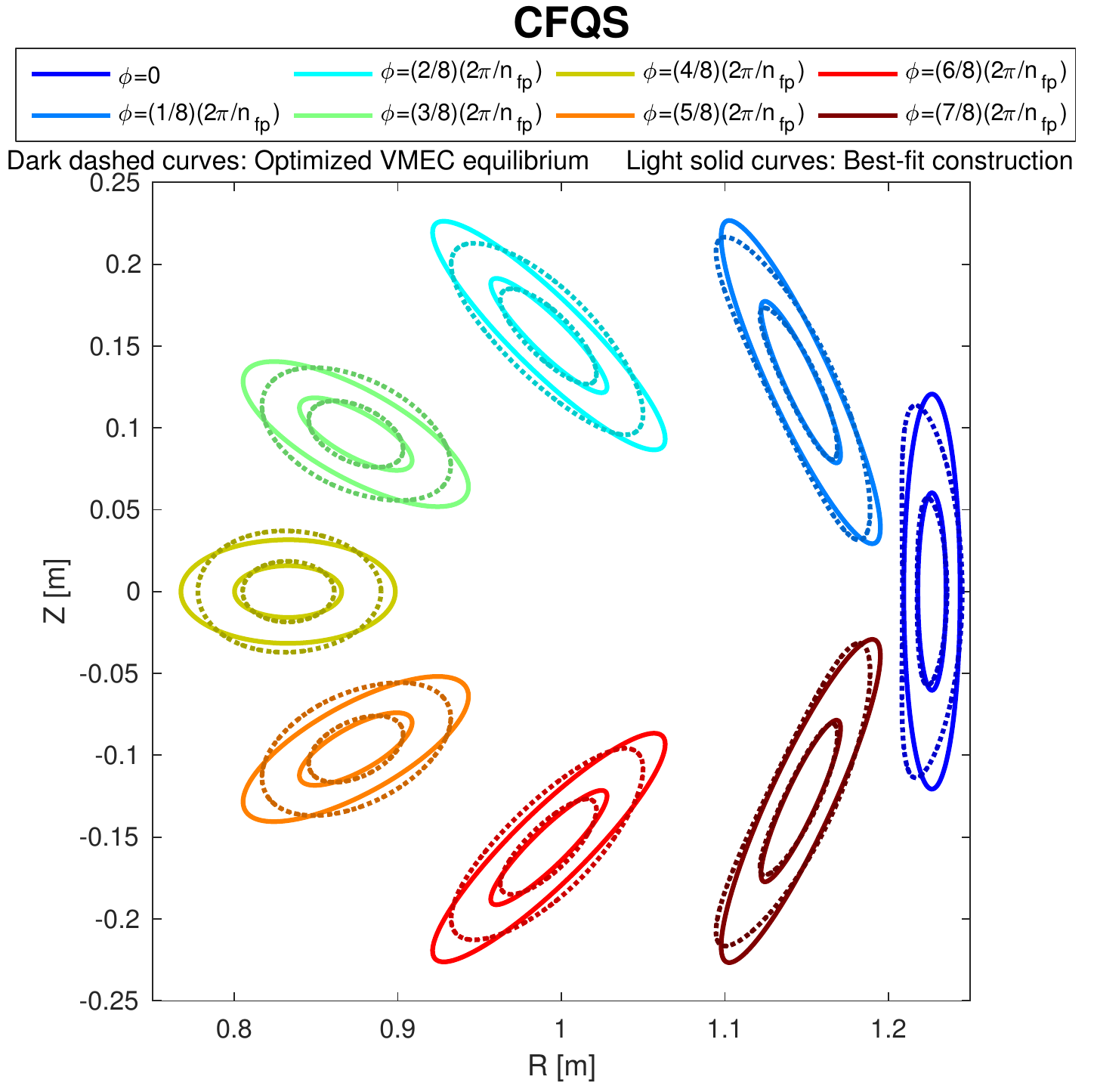}
  \caption{
Shapes of the magnetic surfaces at $\rho=\sqrt{s}=0.1$ and $0.2$ for CFQS \cite{CFQSShimizu,CFQSLiu} (dark dashed curves) at 8 toroidal locations, and the same surfaces of the corresponding direct-construction quasisymmetric configuration (solid curves).
  }
\label{fig:CFQS}
\end{figure}

\begin{figure}
  \centering
  \includegraphics[width=\figurewidth]{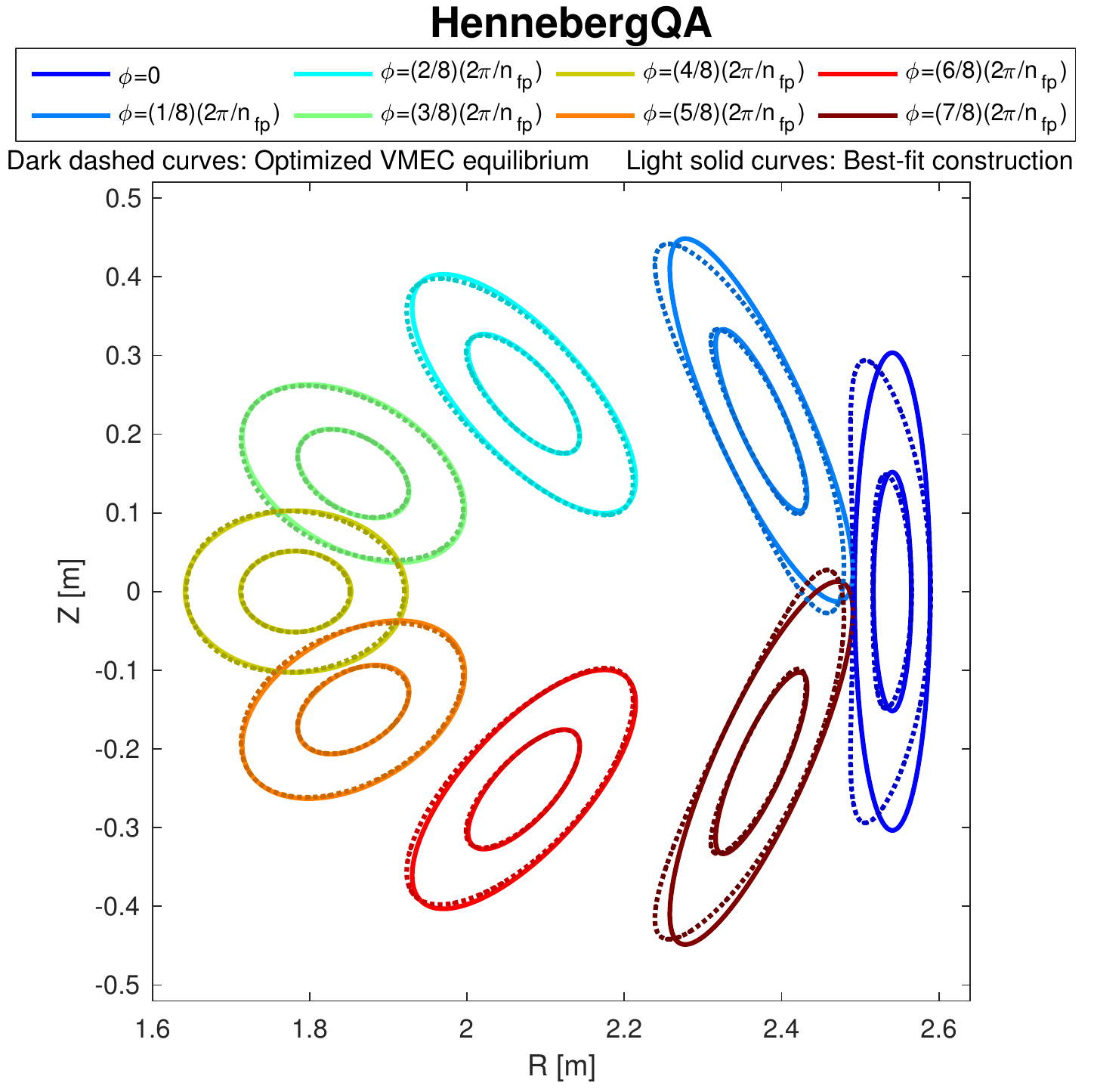}
  \caption{
Shapes of the magnetic surfaces at $\rho=\sqrt{s}=0.1$ and $0.2$ for the configuration of \cite{Henneberg} (dark dashed curves) at 8 toroidal locations, and the same surfaces of the corresponding direct-construction quasisymmetric configuration (solid curves).
  }
\label{fig:Henneberg}
\end{figure}

The fit is carried out for the four quasi-helically symmetric configurations and six quasi-axisymmetric configurations listed in table \ref{tab:configurations}, with the fits shown in 
figures \ref{fig:NuhrenbergZille}-\ref{fig:Henneberg}.
Figure \ref{fig:NuhrenbergZille} shows the earliest quasisymmetric configuration reported, the quasi-helical configuration
from figure 1 of \cite{NuhrenbergZille}.
This configuration contains no pressure or current, has the highest aspect ratio of the configurations considered ($A=12$), and has the highest number of field periods among the configurations ($n_{fp}=6$).
Figure \ref{fig:HSX} shows the only quasisymmetric experiment that has operated to date, the Helically Symmetric eXperiment (HSX)
\cite{HSX}.The quasi-helically symmetric configuration from section 2 of \cite{KuBoozerQHS}
is shown in figure \ref{fig:KuQHS48}, referred to here as KuQHS48 since $n_{fp}=4$ and $A\approx 8$. In contrast to the previous configurations, KuQHS48 has a substantial normalized plasma pressure: $\beta=$4\%.
(All $\beta$ values reported in this paper correspond to VMEC's spatially averaged quantity `betatotal'.)
Figure \ref{fig:DrevlakQH} shows a quasi-helically symmetric configuration reported in Ref \cite{DrevlakKyoto}, developed
recently at the Max Planck Institute for Plasma Physics in Greifswald, Germany.
Moving to quasi-axisymmetric configurations, figure \ref{fig:NCSX} shows the
National Compact Stellarator eXperiment (NCSX) (configuration LI383) \cite{NCSX}.
Figure \ref{fig:ARIESCS} shows the ARIES-CS reactor design (configuration N3ARE) \cite{ARIESCS}.
Figure \ref{fig:QAS2} shows a configuration referred to as QAS2 \cite{GarabedianPNAS,GarabedianNIST}, which has the lowest aspect ratio of the configurations considered: $A=2.6$.
(The $\iota$ profile for QAS2 given in \cite{GarabedianNIST} implies a nonzero current on axis;  we use a configuration with slightly shifted current profile in which the on-axis current density vanishes.)
The ESTELL configuration \cite{ESTELL}, also discussed in section 6 of \cite{ROSE},
is shown in figure \ref{fig:ESTELL}.
A free-boundary $\beta=0$ configuration of the Chinese First Quasi-axisymmetric Stellarator (CFQS) \cite{CFQSShimizu,CFQSLiu} is shown in figure \ref{fig:CFQS}.
Finally, figure \ref{fig:Henneberg} shows the configuration of Ref \cite{Henneberg}.

In all cases, the $\etabar$ fit is performed using the surface at normalized minor radius $\rho=\sqrt{s}=0.1$ from the optimized configuration.
Here, $s=\psi / \psi_{edge}$ is the  toroidal flux normalized to its value at the VMEC boundary.
The figures show the surface shapes at 8 equally spaced toroidal locations. 
For most configurations, the $\rho=0.1$ and $0.2$ surfaces are shown; however for NCSX and ARIES-CS
only the $\rho=0.1$ surfaces are displayed for clarity. 
The corresponding on-axis values of rotational transform $\iota$ from the optimized and directly constructed configurations are compared in
figure \ref{fig:iota_comparison}. The same values are displayed in table \ref{tab:configurations},
along with the best-fit values of $\etabar$.
\changed{
To make the latter independent of device size, $\etabar$ here is normalized by the average major radius of the axis $R_0^{av}$, with the average taken with respect to arclength $\ell$: $R_0^{av} = (\int d\ell\, R)/\int d\ell$.
}

\begin{table}
\caption{\label{tab:configurations}
Quasisymmetric configurations considered in this study.
}
\begin{indented}
\item[]\begin{tabular}{@{}lllllll}
\br
&&Field&Plasma&$\iota$ from&&\\
&Aspect&periods&pressure&optimized&$\iota$ from&Best-fit \\
Configuration&ratio $A$&$n_{fp}$&$\beta$&configuration&construction&$\etabar R_0^{av}$\\
\mr
N\"{u}hrenberg-Zille &12&6& 0\%&1.42 & 1.42 & 1.71\\
HSX&10&4&0\%&1.05&1.06&1.64\\
KuQHS48&8.1&4&4\%&1.29&1.27&1.18\\
Drevlak \cite{DrevlakKyoto}&8.6&5&4\%&1.50&1.50&1.67\\
NCSX&4.4&3&4\%&0.392&0.408&0.594\\
ARIES-CS&4.5&3&4\%&0.412&0.498&0.596\\
QAS2&2.6&2&3\%&0.260&0.267&0.750\\
ESTELL&5.3&2&0\%&0.202&0.201&0.789\\
CFQS&4.3&2&0\%&0.382&0.515&0.586\\
Henneberg \cite{Henneberg}&3.4&2&3\%&0.317&0.314&0.637\\
\br
\end{tabular}
\end{indented}
\end{table}

\section{Discussion and conclusions}

\begin{figure}
  \centering
  \includegraphics[width=3in]{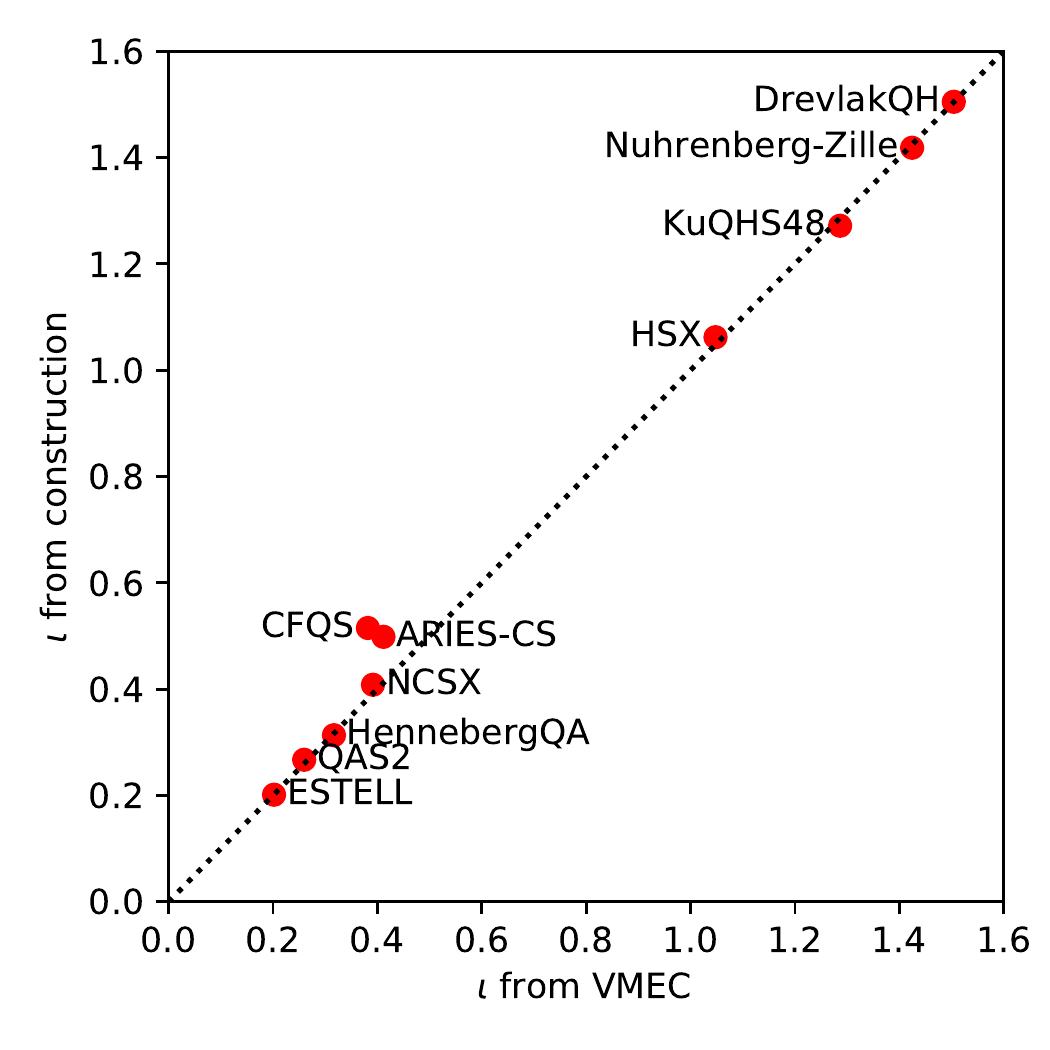}
  \caption{
The on-axis rotational transform computed by VMEC for the optimized configurations generally agrees with the on-axis
transform computed by the Garren-Boozer direct construction.
  }
\label{fig:iota_comparison}
\end{figure}

Figure \ref{fig:iota_comparison} indicates that the Garren-Boozer construction gives an accurate
model for the rotational transform of optimized quasisymmetric configurations. The only  
configurations for which the disagreement exceeds 5\% are ARIES-CS and CFQS.
Shortly, we will show these two exceptions can be understood as the result of relatively large departures from quasisymmetry near the axis in the optimized configurations.

\begin{figure}
  \centering
  \includegraphics[width=3in]{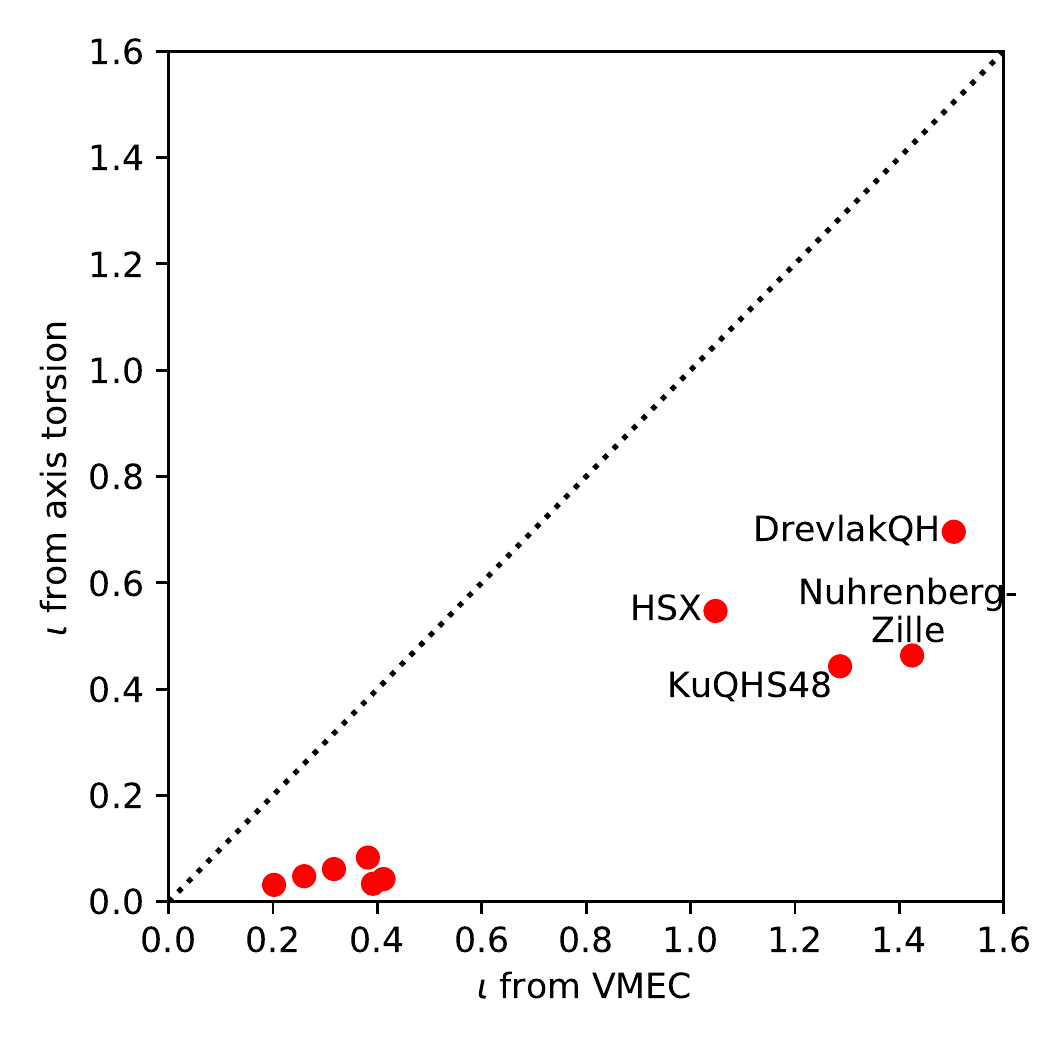}
  \caption{
  \changed{
For each of the 10 optimized configurations, the on-axis rotational transform 
is significantly greater than the contribution from axis torsion. Thus the good agreement
in figure (\ref{fig:iota_comparison}) depends critically on the 
Garren-Boozer construction's ability to predict the elongation and its toroidally rotating orientation. 
Labels are omitted for the six quasi-axisymmetric configurations for legibility.
  }}
\label{fig:iotaFromTorsion}
\end{figure}

\changed{
One might wonder if the agreement in figure (\ref{fig:iota_comparison}) merely comes from the
fact that the axis shapes from the optimized
configurations are used for the construction, since axis torsion gives rise to rotational transform
\cite{Mercier,HelanderReview}.
This hypothesis is seen to be false in figure (\ref{fig:iotaFromTorsion}), which shows that
the rotational transform in the ten configurations is significantly larger than the contribution from integrated axis torsion (eq (44) in \cite{HelanderReview} with $J=0$ and $\eta=0$). 
Evidently the configurations all have significant rotational transform from the other available mechanism: elongation with an orientation angle that varies toroidally \cite{Mercier,HelanderReview}. (The third mechanism to generate on-axis transform, on-axis toroidal current, is absent in all these configurations.) Therefore the good agreement in figure (\ref{fig:iota_comparison})
is meaningful, depending on the ability of the Garren-Boozer construction to match the elongation and orientation angle of the elliptical surfaces.
}

\begin{figure}
  \centering
  \includegraphics[width=5.5in]{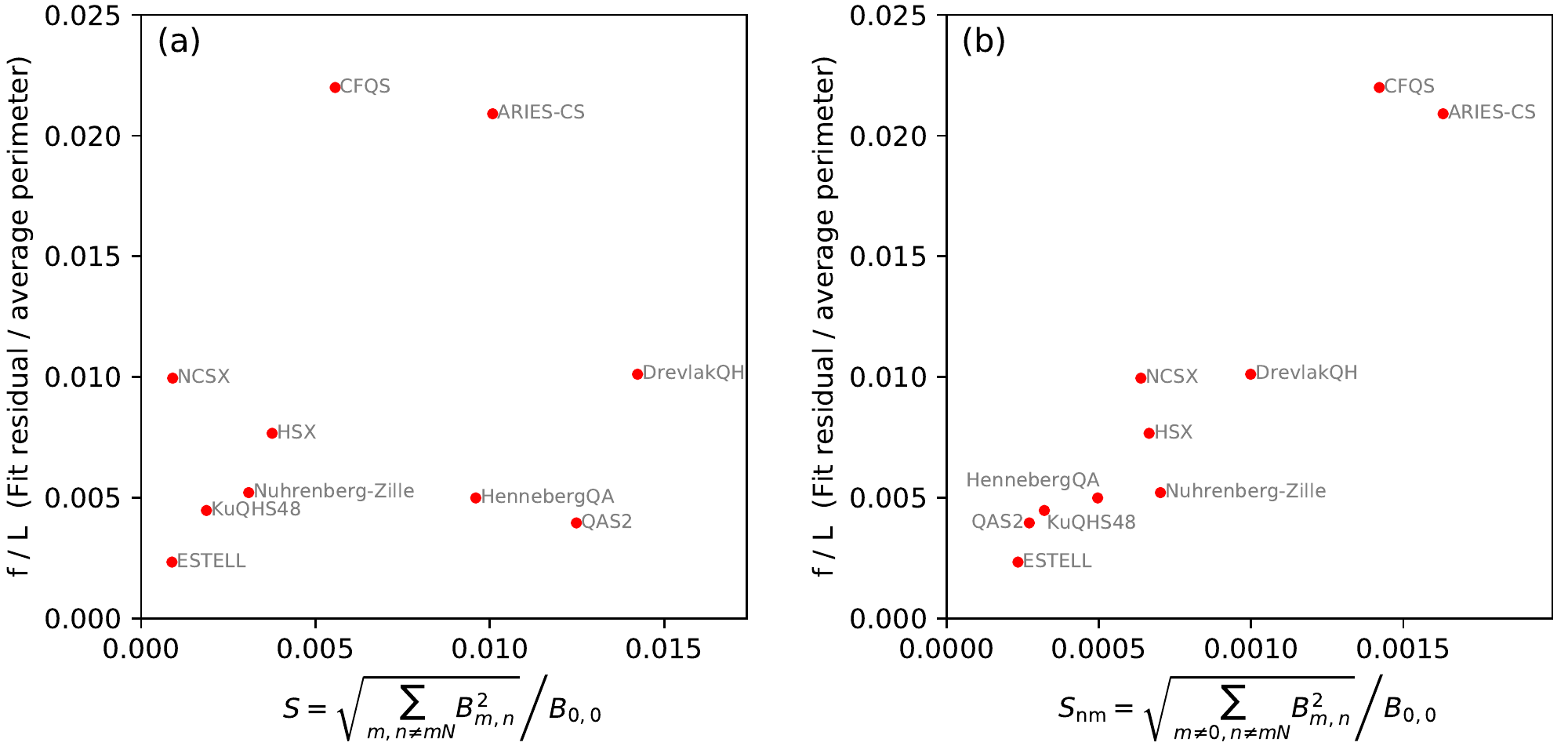}
  \caption{
  \changed{
The difference in the flux surface shapes between an optimized configuration and the best-fit direct-construction configuration ($f/L$) 
is plotted against two measures of symmetry-breaking.
The mirror modes ($m=0$) are included in the $x$-coordinate in (a) and excluded in (b).
All quantities refer to  $\rho=0.1$.
  }}
\label{fig:errors}
\end{figure}

In all cases, except perhaps ARIES-CS and CFQS,  figures \ref{fig:NuhrenbergZille}-\ref{fig:Henneberg} show very close agreement between the shapes of the optimized configurations and
the corresponding constructed ones.
Unsurprisingly the agreement is better at $\rho=0.1$ than at $\rho=0.2$; by $\rho=0.2$ the optimized configurations have some noticeable triangularity (particularly at $\phi=0$) which cannot be represented in the construction.
The fact that near-axis flux surfaces are elliptical in both the optimized and direct-construction configurations 
is not surprising,
since this shape follows automatically from regularity considerations.
The fact that the ellipse centers coincide between the optimized configurations and the analogous constructed configurations is also
no surprise, since the axis shape from the optimized configurations was used as input for the construction.
However, the elongation of the ellipses at each $\phi$ and the angle in the $R$-$z$ plane by which the ellipses are rotated
at each $\phi$ is significant.
The agreement of the elongation and rotation of the flux surfaces between the optimized and constructed configurations demonstrates that the Garren-Boozer model is an accurate model
for the core of realistic quasisymmetric configurations.
Even for ARIES-CS, and CFQS, where the construction tends to over-predict the elongation,
the construction correctly predicts the elongation to be highest at $\phi=0$, and correctly models the trend in the angle of the ellipse major axis versus $\phi$.

The comparison between surface shapes for the optimized configurations and the Garren-Boozer model can
be made more quantitative, and we can 
\changed{explore how}
the departure from quasisymmetry
in optimized configurations correlates with the departure from the Garren-Boozer model, as follows. The departure of an optimized configuration from the best-fit constructed configuration (for a given surface $\rho$)
may be measured by $f/L$, where $f$ is the fit residual in (\ref{eq:residual}), and $L$ is the toroidally
averaged perimeter of the surface cross-sections in the $R$-$z$ plane.
In contrast to $f$, the dimensionless ratio $f/L$ is independent of the overall scale of the configuration.
Next, the departure of an optimized configuration from quasisymmetry (for a given surface $\rho$)
can be measured 
\changed{using the  amplitudes $B_{m,n}(r)$ in the Fourier series $B(r,\theta,\varphi) = \sum_{m,n} B_{m,n}(r) \cos(m\theta-n\varphi)$.
A plausible measure of the departure from quasisymmetry is then
\begin{align}
S=\frac{1}{B_{0,0}} \sqrt{\sum_{m,n\ne mN} B_{m,n}^2}.
\end{align}
Figure \ref{fig:errors}.a shows the fit residual $f/L$ plotted against $S$.
There is no clear correlation. 
Evidently, $S$ is a poor predictor of the quality of fit
to the construction.
However, in figure \ref{fig:errors}.b, the same residual $f/L$ is plotted against a modified measure of the departure from
quasisymmetry, in which the mirror modes (those with $m=0$) are excluded:
\begin{align}
S_\mathrm{nm}=\frac{1}{B_{0,0}} \sqrt{\sum_{m\ne 0,n\ne mN} B_{m,n}^2}.
\end{align}
There is clear correlation between $S_\mathrm{nm}$ and $f/L$.
Figure \ref{fig:errors}.b also shows the two configurations for which the match to $\iota$ is poor in figure \ref{fig:iota_comparison}, ARIES-CS and CFQS, have significantly larger $S_\mathrm{nm}$ than the other configurations.
Comparing panels (a)-(b), it can be seen that three devices (DrevlakQH, QAS2, and HennebergQA) have relatively large mirror modes and yet fit the Garren-Boozer construction well. We can conclude that the fit to the Garren-Boozer model is most sensitive to the nonzero-$m$ symmetry-breaking modes, and not to the mirror modes.}

\changed{
This phenomenon can be understood as follows. A mirror mode arises when the cross-sectional area of the flux surface varies toroidally. The mirror mode amplitude is $<0.015 B_{0,0}$ in all ten configurations considered here,
so the toroidal variation of the cross-sectional surface area is minor. However, the nonzero-$m$ modes are related to the elongation and orientation of the flux surfaces at leading order. This relation can be seen in eq (1)-(2) in \cite{GB1}, which show
the $m=1$ terms in $B_{m,n}$ control the leading-order size of the flux surface in the direction of axis curvature. Hence the $m=1$ terms affect the leading-order elongation, so these modes have a significant effect on the match to the Garren-Boozer geometry. 
There is no correlation in figure \ref{fig:errors}.a because $S$ near the axis is dominated by the mirror modes, the only modes that can have nonzero amplitude on axis, but as explained above the Garren-Boozer model is not sensitive to the mirror mode amplitude.
}

\changed{
Considering that $S_\mathrm{nm} \ll 1$ for all cases in figure \ref{fig:errors}.b, even for CFQS and ARIES-CS where the Garren-Boozer model is not a perfect fit, quasisymmetry must hold near the axis
to high accuracy}
for the construction to fit well. 
Even in a nominally quasisymmetric configuration, 
\changed{
$S_\mathrm{nm}$ might not be extremely small} for several reasons.
Since quasisymmetry cannot be achieved perfectly throughout a volume \cite{GB1,GB2}, it may be preferable to achieve the best quasisymmetry off-axis \cite{Henneberg}, leaving some symmetry-breaking on axis. Some breaking of quasisymmetry may actually benefit confinement: the significantly larger symmetry-breaking in ARIES-CS compared to NCSX was the result of optimization for fast-particle confinement, as discussed in \cite{MynickBoozerKu}. 
Aside from the accuracy of the fit here, both ARIES-CS and CFQS display good quasisymmetry 
in the sense that
the toroidal mode $(m,n)=(1,0)$ dominates across the confinement region.
Evidently quasisymmetry must be achieved to  high precision near the axis in order for the Garren-Boozer construction to be accurate. Nonetheless most optimized configurations have sufficiently good near-axis quasisymmetry that the construction fits well in most cases in practice.

In summary, it has been shown here that the Garren-Boozer model for direct construction of quasisymmetric stellarators can accurately represent the core of most quasisymmetric configurations
in the literature that were obtained using optimization. 
This agreement holds for both quasi-axisymmetric and quasi-helically symmetric configurations, designed by multiple independent researchers using different optimization codes, at a range of aspect ratios, $\beta$ values, and numbers of field periods.
The construction can be fit to a given configuration by tuning a single parameter ($\bar{\eta}$).
The quality of the fit improves as the departure from quasisymmetry in the optimized configuration decreases. Even if a configuration has good quasisymmetry on most flux surfaces,
the elongation and $\iota$ from the construction becomes a poor fit if the near-axis symmetry-breaking 
\changed{
measure $S_\mathrm{nm}(\rho=0.1)$ approaches $\sim0.0015$.}
Since the Garren-Boozer model provides a reasonably accurate analytic description
of the core of quasisymmetric configurations, it can be trusted to provide analytic insight. Furthermore, since the construction can be evaluated at much lower computational cost than conventional optimization, it is feasible to perform wide high-resolution numerical surveys of the space of possible quasisymmetric configurations.
Both of these applications of the Garren-Boozer model will be explored in future work.


\ack
This study was made possible by contributions of magnetic configurations from many people.
Thank you to J Schmitt for providing the HSX configuration, 
N Pomphrey for providing the NCSX, ARIES-CS, and KuQHS48 configurations, G McFadden for providing the QAS2 configuration, 
S Okamura for providing the CFQS configuration,
M Drevlak for providing ESTELL and the configuration of \cite{DrevlakKyoto}, and S Henneberg for providing the configuration of \cite{Henneberg}.
Discussions with M Zarnstorff, N Pomphrey, and H Mynick regarding ARIES-CS are also acknowledged.
The manuscript was improved thanks to feedback from A Geraldini, E Paul, \changed{and J Loizu.}
This work was supported by the
U.S. Department of Energy, Office of Science, Office of Fusion Energy Science,
under award number DE-FG02-93ER54197.
This work was also supported by a grant from the Simons Foundation (560651, ML).


\appendix


\section*{References}

\bibliographystyle{unsrt}
\bibliography{quasisymmetricShapesMatchAnalyticModel}

\end{document}